\documentclass[prl,twocolumn,epsfig,aps]{revtex4}
\usepackage{epsfig}

\begin{document}

\title{Role of calcium and noise in the persistent activity of an
  isolated neuron}

\author{Simona Cocco}
\affiliation{CNRS-Lab. 
Physique Statistique de l'ENS, 24 rue Lhomond, 75005 Paris, France.
\\CNRS-Lab. Dynamique Fluides Complexes, 3 rue de
  l'Universit\'e, Strasbourg, France }
\date{\today}
\begin{abstract}
The activity of an isolated and auto-connected neuron is studied
using Hodgkin--Huxley and Integrate-and-Fire frameworks.
Main ingredients of the modeling are the  auto-stimulating autaptic 
current observed in experiments, with a spontaneous synaptic liberation
noise and a calcium--dependent negative feedback mechanism. 
The distributions of inter-spikes intervals  and
burst durations are analytically calculated, and show a good agreement
with experimental data. 
\end{abstract}
\maketitle


Understanding the mechanisms responsible for persistent activity and
rhythm settling is of central importance in neuroscience. The
persistence of the activity is the neuronal basis of the working
memory \cite{Hebb} and brains rhythms ranging from about 0.1 to 200 Hz
have been recorded in sleep, waking and pathological states
\cite{Wan03}. Such behaviors are usually network properties; for
instance oscillations are possibly built on the existence of different
e.g. inhibitory and excitatory populations of neurons \cite{Mcc00}. From a
theoretical point of view, the large number of neurons in networks
allows the use of self-consistent (mean-field) methods to determine
properties as the average spike emission frequency
\cite{Wan99,Tab00,Seu00,Bru04,Han01}. It was in particular found 
that a basis for persistent activity (with firing rate $\sim 10-50$ Hz) 
is the ability of NMDA synaptic channels to integrate afferent inputs
with a slow decay time constant ($\sim 0.1$ s) \cite{Wan99}. 

\begin{figure}
\begin{center}
\hskip -6cm
\epsfig{file=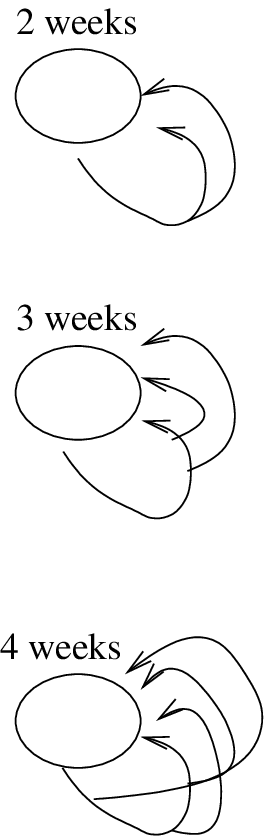,angle=0,width=2cm,height=5.2cm}
\vskip -6.cm
\hskip 2 cm
\epsfig{file=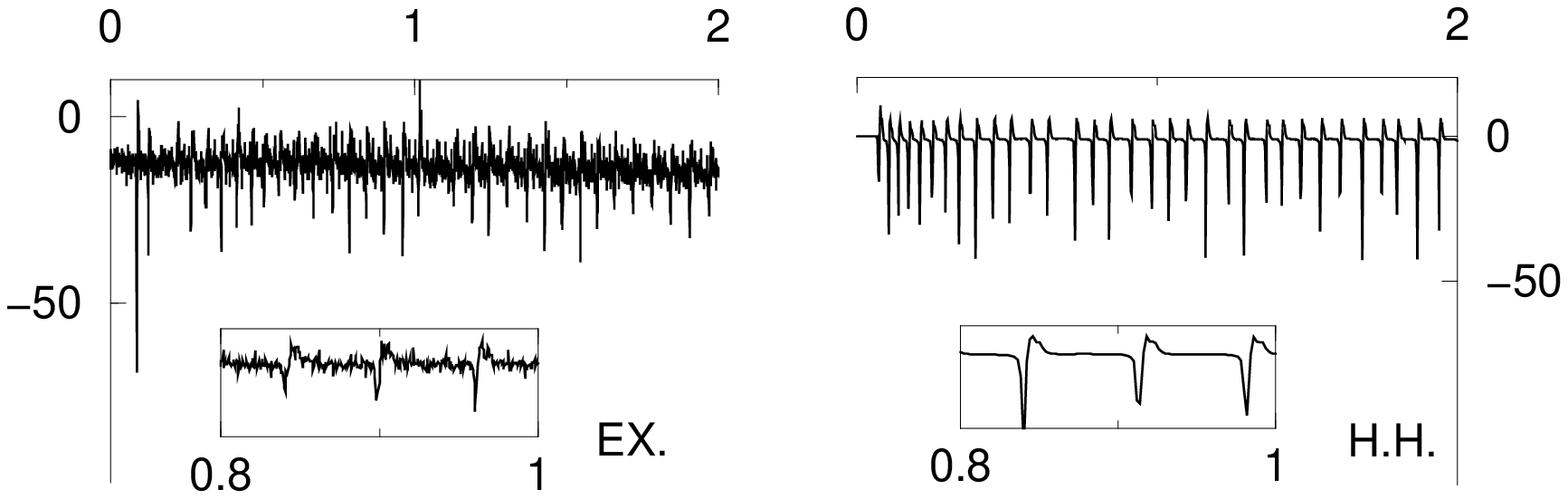,angle=-90,width=6cm}
\\ \hskip 2 cm
\epsfig{file=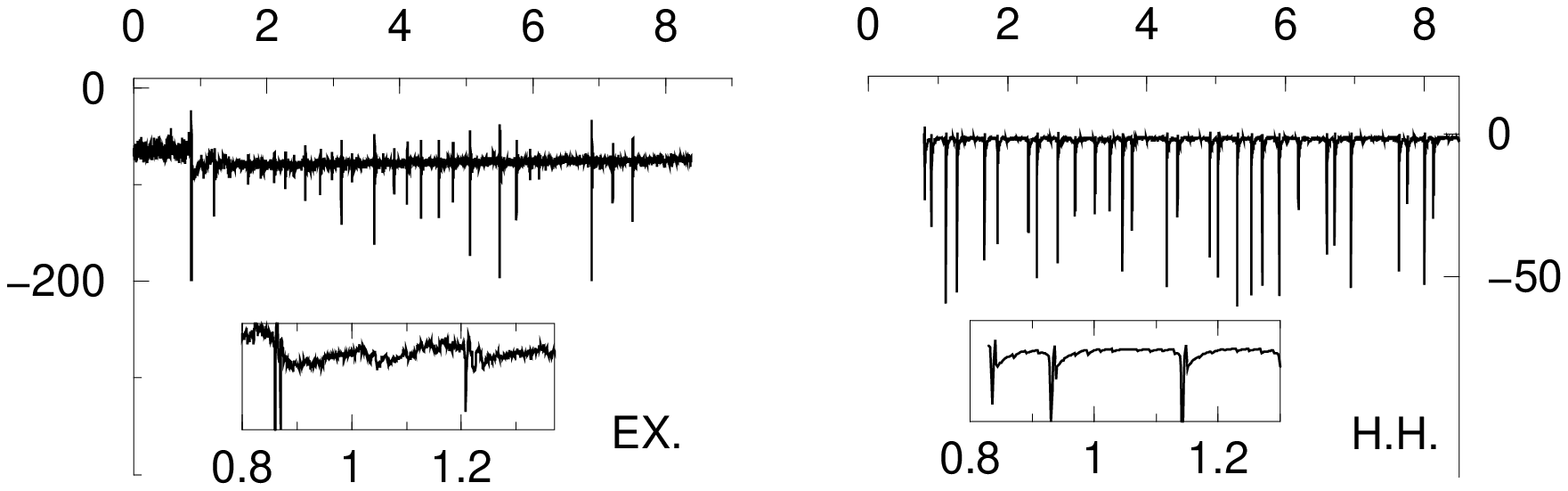,angle=-90,width=6cm}
\\ \hskip 2 cm
\epsfig{file=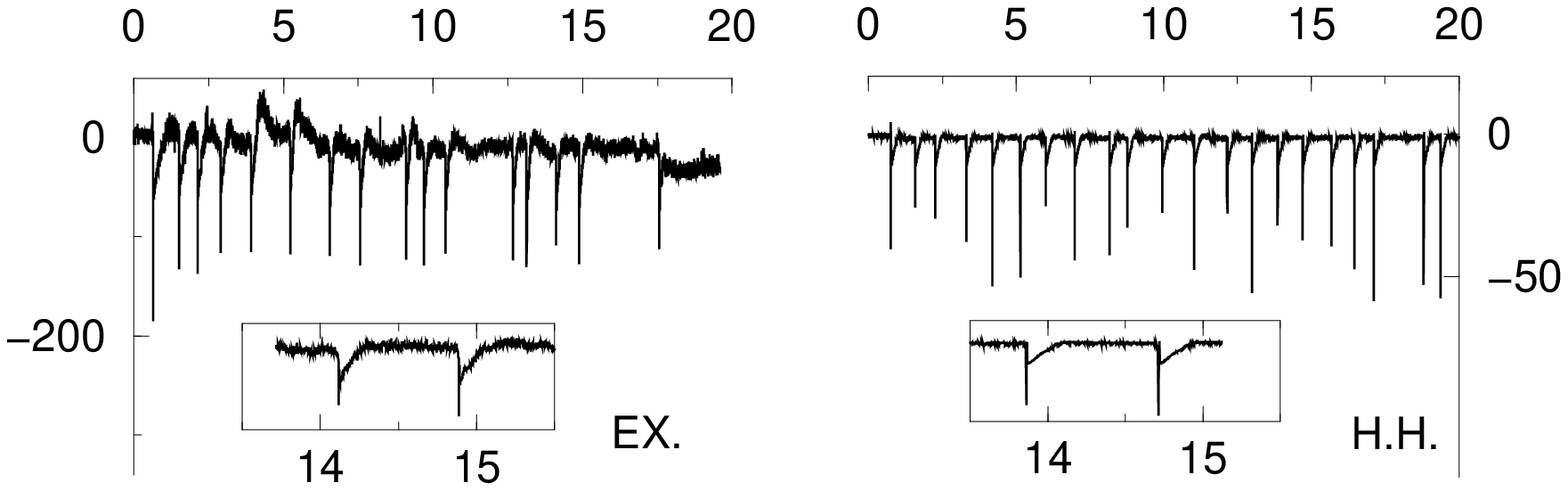,angle=-90,width=6cm}
\caption{Sketch of the autaptic system (left panel) showing the
increase in synaptic connections with age (2, 3 and 4 weeks from top
to down) and experimental (EX, middle) \& theoretical (HH model,
right) current time traces. Units are pA (EX) \& $\mu$A/cm$^2$ (HH)
vs.  seconds. Experimental data show bursts of activity for 3
representative neurons among 20 records.  HH simulation show current
dynamics for bursts of equivalent durations.  Insets magnify two
consecutive spikes chosen in the bursts, with typical ISI of 0.05, 0.2
and 1 s for 2, 3 and 4 weeks respectively.  The autaptic AMPA current
(negative {\em i.e.} inward, immediately after the spike) is very small
and not visible at 2 weeks, large and observable at 3 weeks, huge at 4
weeks.}
\label{curr}
\end{center}
\vskip -1cm
\end{figure}

Interestingly, persistence and rhythm settling have also been
experimentally observed in systems made of an isolated and
auto-connected excitatory neuron (autapse) \cite{noi}.
Autapse are produced {\em in vitro} by grown excitatory neurons,
extracted from rat embryos hippocampal, on coverslips
\cite{Wya02,noi,Bek}. Neurons normally develop up to 5 weeks and establish
connections with themselves when no other neuron is nearby. The number
of auto-connections increases with the age, as sketched in
Fig.\ref{curr}.  Patch pipettes allows both electrical recording of
the neuronal activity and current injection to trigger spikes.
  After a spike has been triggered, more than 2 week old
autapses carry on spiking in a whole burst of activity.  Records show
that both the time interval between successive spikes (ISI) and the
duration of the burst (BD) fluctuate.  Surprisingly, while the number
of auto-connections increase with the age, the average spike frequency
decreases (around 20, 5, 1 Hz for 2, 3, 4 week neurons respectively),
see Fig.\ref{curr}.  The neuron therefore exhibits a negative
rate-control feedback mechanism preventing runaway excitations due to
the strong, and growing with the age, positive auto-stimulating
current. All those experimental results are reported in \cite{noi}.

This letter presents a theoretical study allowing a quantitative
interpretation of the autapse activity.  Persistence is due to the
interplay of two currents evidenced in experiments \cite{noi}: a small
and slow postsynaptic component and a random spontaneous synaptic
liberation. To model rhythm settling, a calcium dependent negative
feedback is introduced. This feedback is at the origin of spike
frequency adaptation under an external current, previously modeled 
\cite{Wan98,Wan01} and experimentally observed in the autaptic system
\cite{Wya02}. ISI and BD distributions are analytically calculated
using an Integrate--and--Fire (IF) model, and compared to experimental
data and numerical predictions from a detailed Hodgkin-Huxley (HH)
model.

\begin{figure}
\begin{center}
\epsfig{file=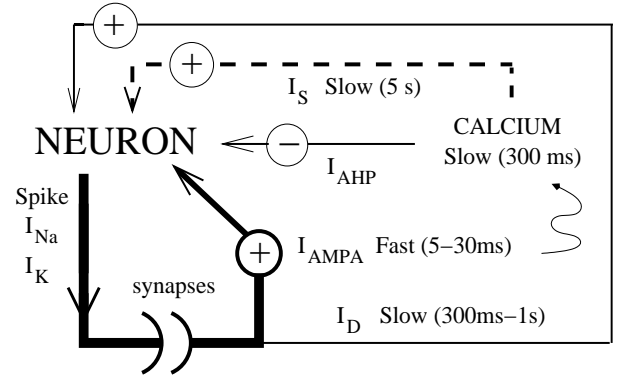,angle=0,width=8cm,height=5cm}
\caption{Currents in the auptatic neuron model, with line widths
proportional to amplitudes. Presynaptic 
currents $I_{Na}$ and $I_K$ accompany a spike.
 Postsynaptic components are, see text: $I_{AMPA}$ (thick line), 
responsible for membrane depolarization and more Ca entering the cell
(wiggly line); $I_D$, a slow depolarizing current (thin line);
$I_S$, due to enhanced random spontaneous miniatures following
Ca release (dashed line). The negative feedback is accounted for by
a potassium hyperpolarizing current, $I_{AHP}$, with Ca-dependent amplitude.}
\label{schema}
\end{center}
\end{figure}

\begin{figure}
\begin{center}
\epsfig{file=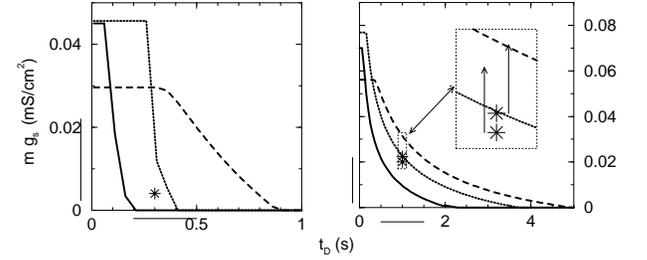,angle=-90,width=8cm}
\caption{Phase diagram for the persistence of activity in $(t_D, m\,g _s)$ 
plane for $g_l=0.1, \theta=-53.5$ mV (left) and 
$g_l=0.2, \theta=-52$ mV (right). Bars symbolize ranges of 
experimentally observed values \cite{noi}.
 Parameters are:
$Ca ^{sp}$=0.043 (2 week, full), 0.115 (3 week, dotted) and 0.55 $\mu$M 
(4 week neuron, dashed line). Activity persists above the lines,
is absent below.   
Left: parameters, indicated by the star in (0.3,0.004) 
with $g_s=0.002, m=2$, for a 2 week cell with stationary ISI of 0.07 s.
Right: stars lies slightly  below 
threshold at (1,0.02) --$g_s=0.0032,m=6.5$ (3 weeks)--
and (1,0.022) --$g_s=0.0032,m=7$ (4 weeks). 
Positive fluctuations of the noise are needed for 
the activity to persist. Inset: magnification of the star--surrounding
region; arrows indicate a positive
standard deviation reaching (3 weeks: narrow line on the left)
or  (4 weeks: bold line on right) crossing the critical line. 
 Average ISI are 0.2, 0.75 s for 3, 4 weeks respectively.} 
\label{fasediag}
\end{center}
\vskip -1cm
\end{figure}

Currents in the autaptic system are represented in Fig~\ref{schema}.
Pre-synaptic (spike) currents are introduced in HH through standard
Sodium and Potassium gating variables \cite{Hod52,nota3}; their
modeling in IF is discussed below. The experimental characterization
of the post-synaptic current \cite{noi} has evidenced three auto-stimulating
components that follow the spike (Fig.~\ref{schema}):

1. A large amplitude excitatory AMPA component, $I_{AMPA}$, entering
just (5 ms) after the spike, and rapidly decaying.  Its amplitude
considerably increases with the age of the neuron, due to the growing
number of connections (Fig.~\ref{curr}). The AMPA current is modeled,
in HH, through 
an effective exponentially decreasing conductance \cite{nota3}.
We stress that the AMPA current arrival 
falls within the neuron refractory period
and thus cannot by itself trigger a new spike; it however
leads to a membrane potential depolarization (up to 0 mV), giving
a bump in the flank of the spike in voltage records \cite{noi}.
The AMPA depolarization has two consequences. It first slows down 
Na and K gating variable resets, increasing the refractory
period. Secondly, it allows more calcium to enter the cell via 
voltage--gated channels, as soon as membrane potential exceeds -20 mV.  
In IF, AMPA,  Na and K currents are altogether
accounted for by the calcium $Ca^{sp}$ entering at each spike, and the 
refractory period $t_r$. These two parameters (which depend on the age
of the autapse) and the threshold $\theta$ for spike firing are
determined from the numerical analysis of HH \cite{io},
see  Fig.~\ref{fasediag} and \cite{nota3}. 

2. A small amplitude and slow decaying component, $I_{D}$, 
due to NMDA and ICAN conductances \cite{noi}. 
$I_D$ depolarizes the membrane potential $V$
during a burst to $\sim -55$~mV; after the burst halts,
$V$ decays to the rest value $V_l=-60$~mV with a time constant $t_D$
ranging from 300 ms in young cells (2 weeks) to 1 s in mature cells (3
and 4 weeks). In both HH and IF, $I_D$ is modeled as an
exponentially decaying current, $I_{D}(t)=I_{D}^0\; e^{-(t-t_i-\delta
  t)/t_D}$, released with a delay $\delta t=5$~ms after the spike
occurring at time $t_i$; the amplitude is set to
$I_{D}^0=2$~mA/cm$^2$, and corresponds to a depolarization of the
membrane potential of 5 mV.

3. A spontaneous and random synaptic release, called miniatures,
$I_S =g_s\,V\,s(t)$.  Both amplitude and frequency of the miniatures
$s(t)$ are stochastic \cite{noi}. The time interval between miniatures
during a burst has been fixed to its average value $\delta _s = 20$
and 10~ms for 2-3 and 4 week neurons respectively.  $\delta _s$
increases to its (much larger) rest value $\sim 0.1$~s in about 5~s
after the end of the burst. Release times are hence discrete and
multiple of $\delta_s$, which makes the IF model mathematically
tractable (taking into account the stochasticity in times between 
miniatures does not significantly affect the outcome \cite{io}). 
The distribution $P$ of the released
amplitude $\sigma$ is Poissonian; the mean $m$
and the conductance $g_s$ are fitted from experiments \cite{nota2},
and given in caption of Fig.~\ref{fasediag}.  After release at time $i\times
\delta _s$, the miniature exponentially decays, $s(t)=(s^{-}+ \sigma
(i))\; e^{- (t-i\,\delta_s)/t _s}$ with $t_s = 5$~ms ($<\delta _s)$
from experiments.  Parameter $s^{-}=m/(e^{\delta _s/t_s}-1)$ is the
average residual amplitude before the release.

In addition, an inhibitory feedback due to the presence of 
a Calcium--dependent After Hyperpolarizing current, $I_{AHP}=g_{AHP}\, 
\frac{Ca}{k_d} \;(V-V_K)$, is introduced. 
Such a component is modeled as in \cite{Wan98} for 
HH and \cite{Wan99,Wan01} for IF. Parameters values are
$g_{AHP}=5$~mS/cm$^2$, $V_K=-80$~ mV, $k_d=30\ \mu$M. 

\begin{figure}
\begin{center}
\epsfig{file=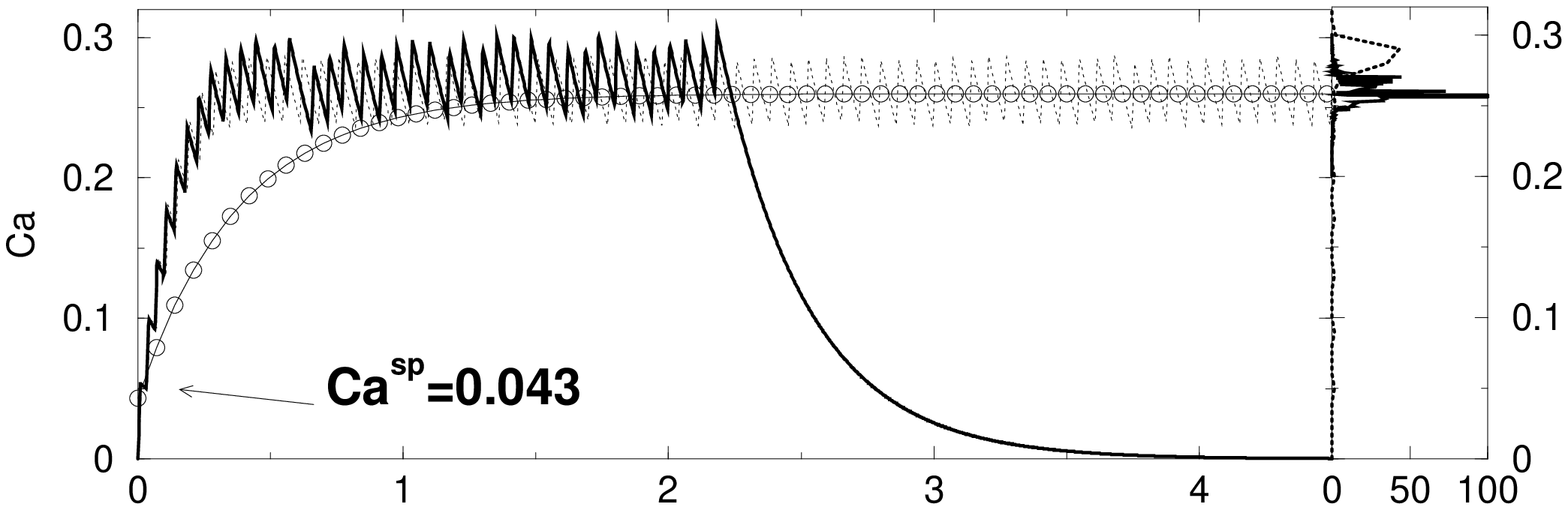,angle=-90,width=5.6cm} 
\epsfig{file=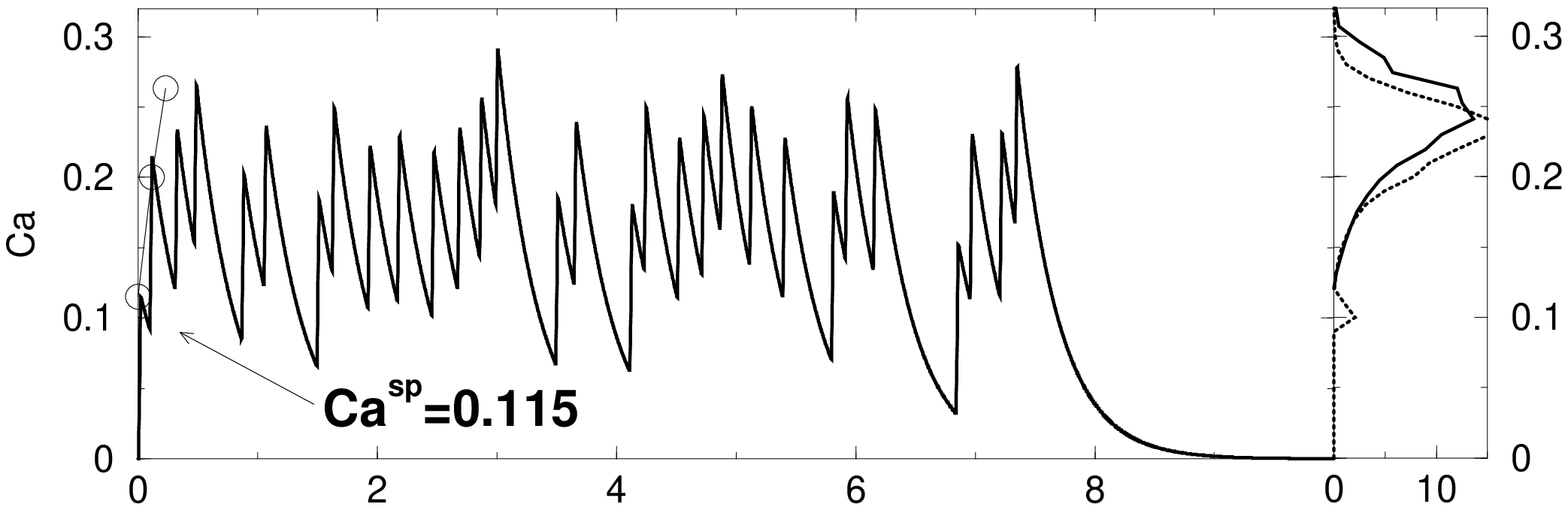,angle=-90,width=5.6cm}
\epsfig{file=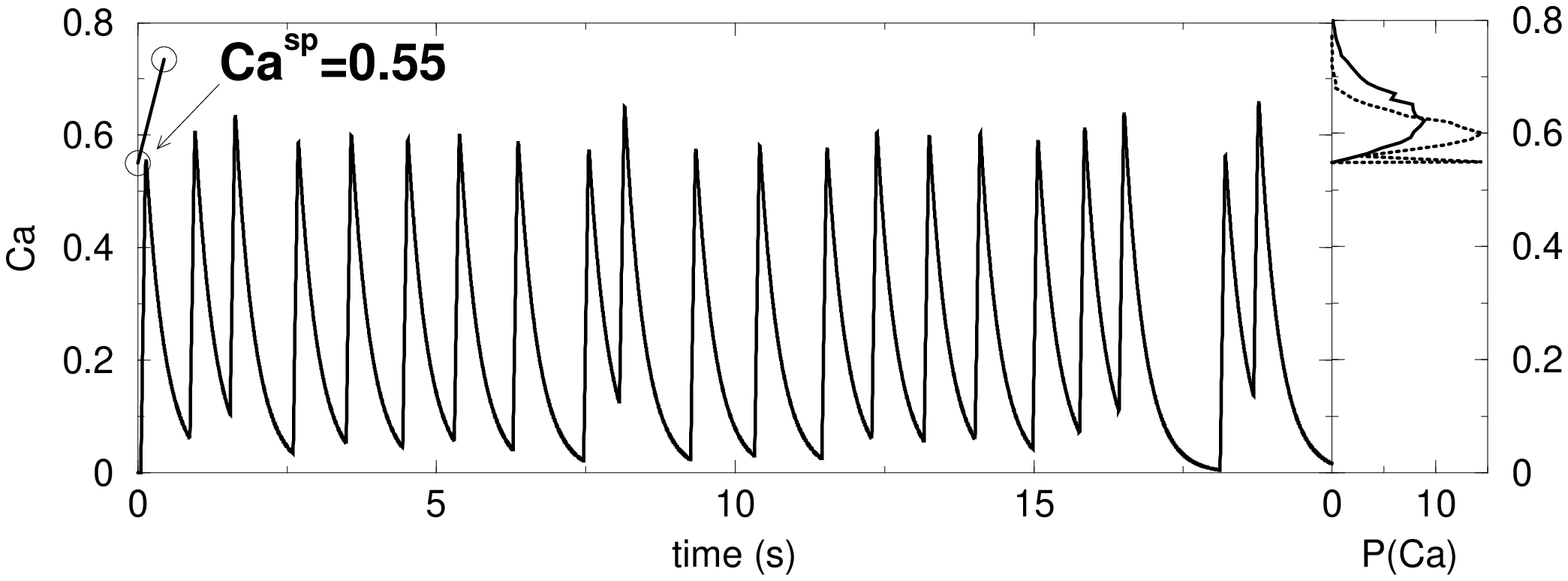,angle=-90,width=5.6cm}
\caption{From top to down, 2, 3 \& 4 week neurons.
Left: Calcium levels vs. time during a burst
 for random --HH, numerical, full line--  and fixed amplitude
--IF, theory, circles-- miniatures. Jumps coincide with spikes.
Right: IF --theory, full-- and HH --numerical, dotted line-- 
distributions of calcium
right after spike emission. Distribution is very peaked 
for the 2 week neuron around $Ca=0.26$ (IF) and $0.28$ (HH), and
much wider due to noise-driven activity in older
cells.}
\label{cadin} 
\end{center}
\vskip -1cm
\end{figure}

 
In the IF model, the dynamics of membrane potential $V$ and calcium
concentration $Ca$ obey
\begin{eqnarray}
\label{IF}
&&C {dV \over dt}= -I_l-I_{AHP}(Ca) +I_{D} - I_S 
\\ && {dCa \over dt}=Ca^{sp} 
\sum_{n} \delta (t-t_n)-\frac{Ca}{t_{Ca}} \label{IFCA}
\end{eqnarray}
where $I_l= g_l\, (V-V_l)$ is the leak current, $C=1$~mF/cm$^2$ 
the capacity, and $t_n$ the emission time of the $n^{th}$ spike. 
A spike is fired when $V>\theta$, and the refractory period $t_r$ is over. 
Large amplitude
synaptic miniatures ($I_S$) that add to the slow component ($I_D$) can
trigger a spike, provided the calcium-based inhibition
($I_{AHP}$) is sufficiently weak. This mechanism is made possible by
the fact that the slow component and high-frequency miniatures persist
over a time period larger than (or, at least, equal to) the calcium 
decay constant $t_{Ca}=0.33$~s.

This few hundred ms time-scale is well separated from the integration
time, $\tau_m=C/g_l=5-10$~ms for $g_l=0.1-0.2$, and miniature decay
time, $t_s=5$~ms . Consequently, once time $t$ after the last
spike emission is expressed in terms of miniature intervals, $t=
i\times \delta _s + \tau$, where $i (\ge 0)$ is integer-valued and
$\tau$ continuously ranges in $[0;\delta _s]$,  $I_{AHP}$ and
$I_D$ do not vary with $\tau$ \cite{Rin89}.  From (\ref{IFCA}), the calcium
level $Ca$ is a function of its value $Ca_-$ right after the previous
spike and the interval $i$: $Ca[Ca_-,i]= Ca_- \; e^{-i \delta _s/
t_{Ca}}$. We then integrate (\ref{IF}) analytically over $\tau$ at
fixed $I_D(i),I_{AHP}(i), \sigma (i)$ and $Ca_-$.  The resulting
expression for the maximum of the membrane potential over $\tau$,
$V_M (i, \sigma (i) , Ca_- )$ \cite{io}, is then compared with
$\theta$. $i$ is increased by one if no spike is fired; otherwise, it
is reset to 0 and the calcium level is increased by $Ca^{sp}$.
Spike emissions (Fig.~\ref{curr}) are therefore reflected by 
jumps in the $Ca$ level, as illustrated in Fig.~\ref{cadin}.

 
Insights about the values of parameters allowing activity to be persistent
can be obtained when considering first non-stochastic 
miniatures with fixed amplitude $m$.
In this case, a stationary dynamics corresponding to a regular burst with
infinite duration can settle down. The
values for the calcium after each spike, $Ca^*$, and ISI, 
$i^*\; \delta _s$, are such that the calcium decrease
between two spikes balances the increase after a spike, 
$Ca^*=Ca^{sp}/(1-e^{-i^*\, \delta_s/t_ {Ca}})$.
Inserting this result into the IF condition 
$V_M( i^*,m,Ca^*)=\theta$ derived above yields 
a self-consistent equation for the stationary ISI value $i^*$. 
The existence or absence of solution to this equation indicates whether 
a persistent activity can be sustained or not \cite{nota5}.
The phase diagram is shown in Fig.~\ref{fasediag}, for two
values of the neuron conductance, $g_l=0.1,0.2$, in the
experimentally measured range. 

\begin{figure}
\begin{center}
\epsfig{file=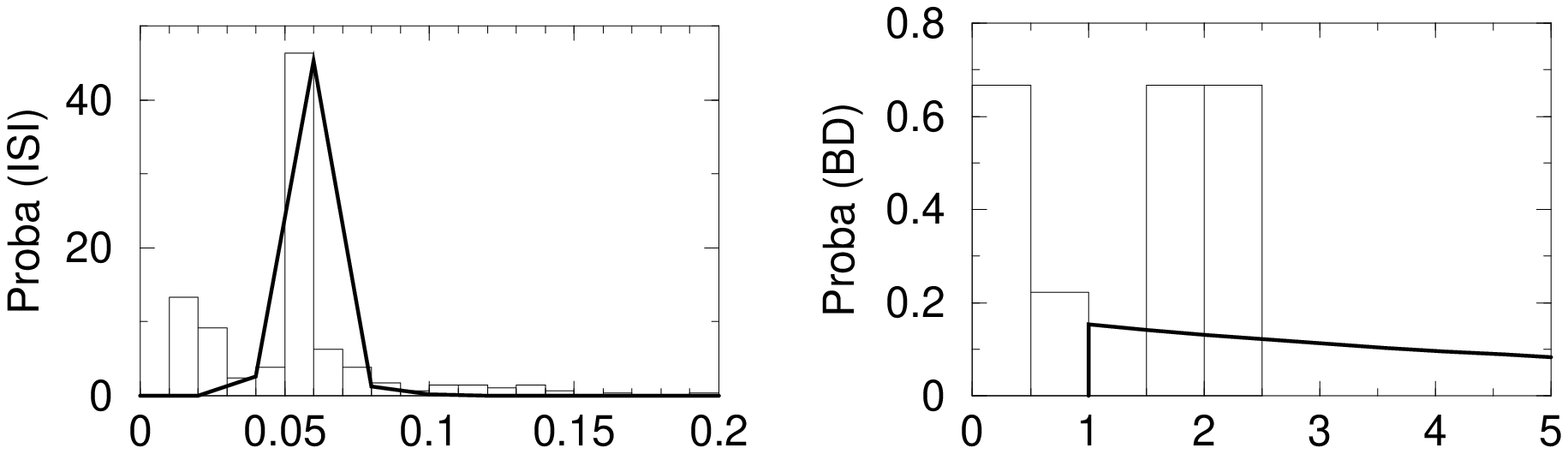,angle=-90,width=6cm} 
\epsfig{file=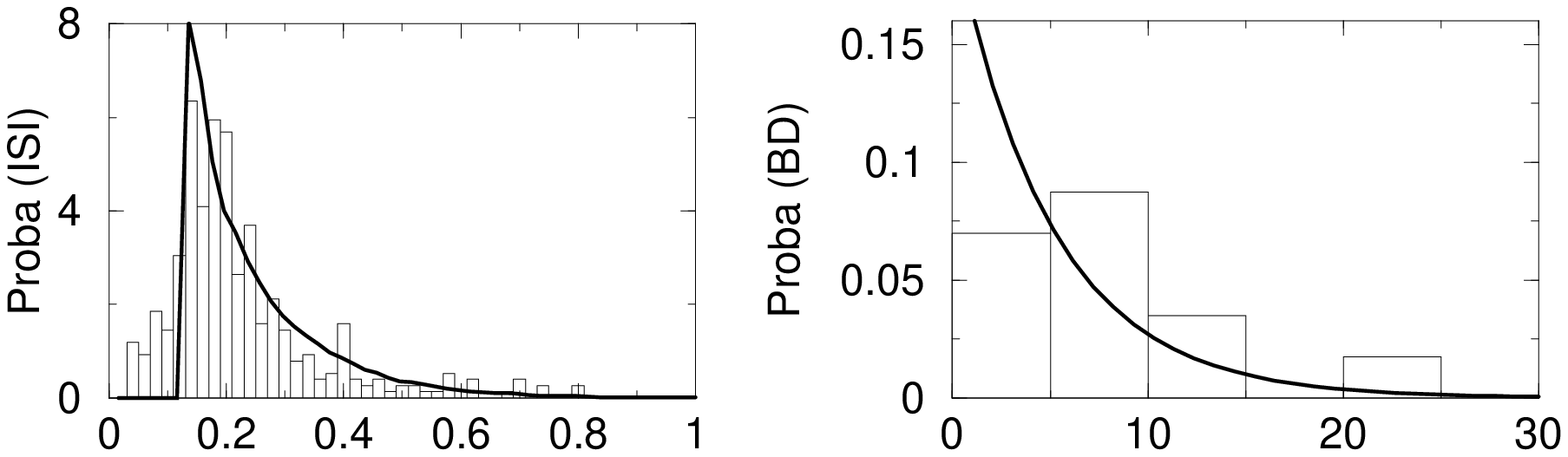,angle=-90,width=6cm}
\epsfig{file=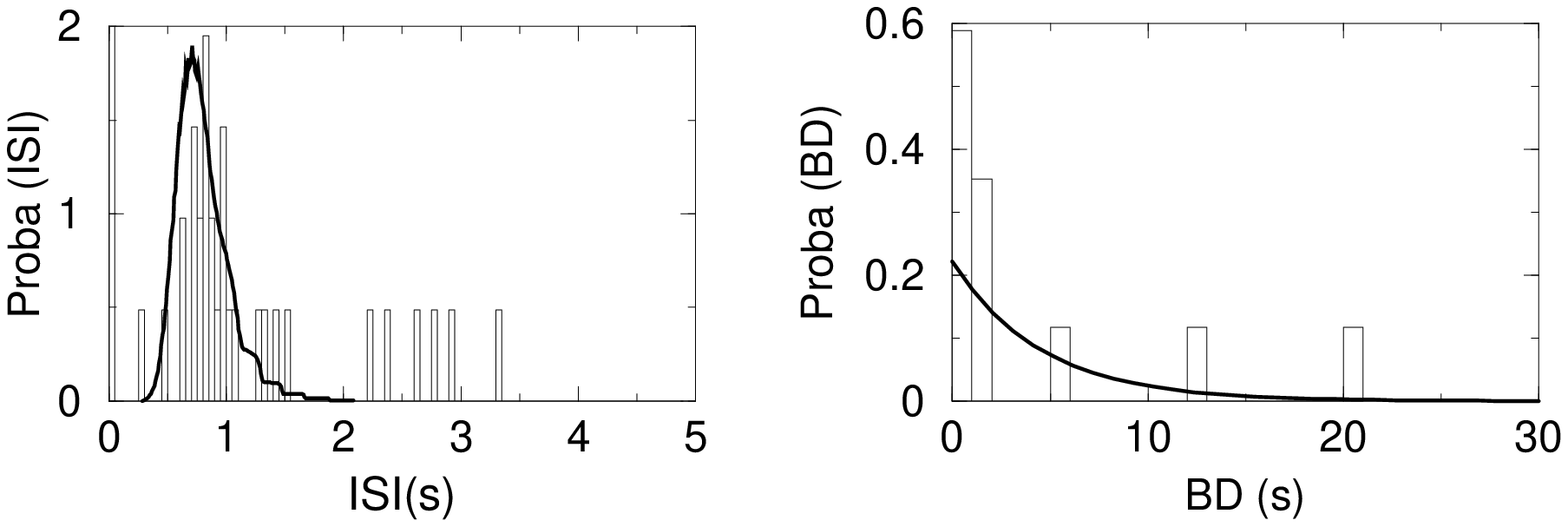,angle=-90,width=6cm}
\caption{ISI (left) and BD (right)
  distributions for 2, 3 \&4 weeks respectively (top to down). 
Histograms are experimental data, full lines are IF theoretical
predictions. HH simulations, coinciding with IF predictions, are 
not shown. }
\label{distributions}
\end{center}
\vskip -1cm
\end{figure}

The presence of noise in miniatures remarkably enriches this
picture. Experimental features of ISI (average value in Fig.~\ref{curr}; 
small fluctuations for 2 week neurons and wide distribution  
for 3-4 weeks in Fig.~\ref{distributions}) are well reproduced
by parameters shown in Fig.~\ref{fasediag}.
The system representative parameters (star) lies above the critical 
persistence activity line for 2 week cells, and  slightly below   
for 3-4 weeks. In the latter case, miniatures showing positive 
fluctuations occasionally make representative parameters cross the phase
border, and the neuron generates spike (noise-driven activity).
On the contrary, in the former case, activity would persist forever 
in the absence of noise.

Finite burst duration results from a two-fold mechanism. On the one hand, 
if  miniature amplitudes $\sigma (i)$ happen to
be lower than the minimal value that can trigger a spike
 in the vicinity of $i^*$, no spike will be
fired. Later on, as the slow component decreases, spike firing becomes
less and less likely.  On the other hand, as a result of positive
fluctuations in the $\sigma$s, a very short ISI may arise, making
calcium increase and further spike firing unlikely. 

The above results are illustrated in Fig.~\ref{cadin} for
both HH/IF models, and noiseless/noisy dynamics.
Note that, for 2 weeks neurons, persistent activity is possible
in the absence of miniatures too ($m=0$) through the slow component
only for 2 week cells. 
A burst, in the noisy dynamics, halts after  a positive
increase in  calcium due to contiguous spikes with short ISI.  
For 3 and 4 week cells the noiseless IF dynamics  
does not reach stationarity, while the noisy dynamics results in
finite and random duration bursts.

To calculate the 
stationary distribution of calcium, ISI, BD in presence of noise, 
we consider the probability $\hat T(i|Ca _-)$ to fire with an ISI 
equal to $i\,\delta _s$
given the calcium value $Ca _-$ after the previous spike:
$\hat T(i | Ca _-)=\prod_{j=1}^{i-1} \bar{p}(j|Ca _-)\,(1-\bar{p}(i|Ca _-))$;
$\bar{p}(j|Ca _-)=\sum_{k=0}^{s_m(j|Ca _-)-1} P(k)$ 
is the probability not to fire at interval $j$  
where  $s _m(j|Ca _-)$ is the minimal amplitude of noise 
able to trigger a spike at interval $j$, $P$ is the
Poissonian with average $m$.
The transition matrix between two successive calcium values read 
$T(Ca|Ca_-)= \sum_i \hat T(i|Ca_-) \delta(Ca - Ca[Ca_-,i]-Ca^{sp})$.
The stationary calcium distribution, $Q(Ca)$, is
the maximal eigenvector of $T$. We have discretized the calcium interval, 
and used Kellogs iterative projection method to diagonalize $T$ with the result
shown in Fig.~\ref{cadin}.  
The ISI distribution can be obtained from the matrix product of $\hat T(i|Ca)$
and $Q$, and is shown in Fig.~\ref{distributions}. It is
very peaked around 0.05 (EX) - 0.06 (IF) s for 2 week cells, and 
spread out for 3 and 4 week neurons, with median 0.2, 0.75
s. respectively,  in good agreement with 
experiments. To calculate the BD distribution, we define the generating
function for the probability of an interval $i\, \delta _s$ between
spikes, $G(x)=\sum_i Q(i) x^i$. The coefficient of $x^k$ in $G(x)^n$, 
denoted by $[x^k]\, G(x)^n$ is the probability that a burst with
$n$ spikes has duration $k\, \delta_s$. Summing over $n$, we get the probability that a
burst has duration $k\, \delta_s$, $[x^k]\;G(x)/(1-G(x))$. 
The resulting BD distribution is shown in Fig.~\ref{distributions}; 
it decays as $e^{-k/k _o}$ where $k_o$ is the root of $G(e^{1/k _o})=1$. 
The agreement with experiments is good, even if 
comparison suffers from limited data (9, 13, 11 bursts for 2, 3, 4 weeks);
it is excellent with HH simulation, which makes the 
analytical study of IF quite attractive despite the approximations done
(discretization of time).

In conclusion, this letter proposes a possible mechanism that accounts for
the persistence of activity in an isolated autapse.
 The slow  current $I_D$ sets the neuron in a depolarized
'up' state, also observed 
 in various oscillatory networks \cite{Mcc00} where it results from
{\em e.g.} afferent inputs from other neurons.
For 2 week neurons this up state can, by itself, sustain activity while, 
for 3 and 4 week, activity requires noisy synaptic liberation.
Noise is at the same time responsible for 
persistent dynamics, robust with respect to 
changes of parameters \cite{nota5}, and the finite duration of bursts.
The noise--driven mechanism for burst halting presented here differs 
from mechanisms based on a slow activity
dependent depression \cite{Tab00} (either due to a modulation
of cellular excitability, or to synaptic depletion \cite{nota6}),
and is supported by the observed large variability
in burst durations. 
In addition, calcium--dependent spike frequency adaptation 
can explain the observed pattern of activity as the increase of 
ISI with the age. Finally, it would be interesting to extend the 
present study to the case of a network composed
of a small number of similar neurons for which experimental data are
available.

Acknowledgments: this work originates from discussions with
C. Wyart, D. Chatenay and R. Monasson. I am particularly grateful
to the latter for critical reading of the manuscript.  

\begin{thebibliography}{99}
\bibitem{nota3} Na and K dynamics 
 parameters are borrowed from \cite{Wan98},  with temperature constant
 $\Phi_m=1,\Phi_h=\Phi_n=0.3$
and conductances $g_{NA}=3 mS/cm^2, g_K=1 mS/cm^2.$ 
The amplitude and time constant for the AMPA component are fixed for
 2, 3, 4
 weeks to 0.1, 3, 20 in $mS/cm^2$ and $ 10, 20, 30 ms$ respectively 
\cite{io}. The refractory period in the IF model is
 $t_r=10, 16, 85$~ms for 2, 3, 4 weeks.
\bibitem {nota2} Mean value and standard deviation of 
 miniatures conductance in active neurons are $g=0.45$ nS, $\sigma=0.2$
 nS respectively \cite{noi}.
Fitting data with a Poissonian distribution \cite{io} and writing 
$g=g_0\,m$, we obtain $m= g^2/\sigma^2=5$, $g_0=\sigma^2/g=.008$ nS,
hence a conductance $g_s=g_0/A=0.002$ mS/cm$^2$ for a cell surface
$A= 4\, 10^{-6}$ cm$^2$.
\bibitem{nota5}
Even if a stationary solution exists, it may be not possible to reach for
dynamical reasons e.g. for young neurons with special choices of 
parameters \cite{io}.
\bibitem{nota6} Synaptic depletion has been neglected here. Note that
experimentally observed autaptic currents are not affected by
repeatedly triggered spikes, except at very high stimulation 
frequency $\simeq 10-20 Hz$ .
\end{thebibliography}
\end{document}